\documentclass[12pt]{article}

\usepackage{mathrsfs}
\usepackage[T1]{fontenc}
\usepackage{mathpazo}
\usepackage{setspace}
\usepackage{amsfonts}
\usepackage{amssymb}
\usepackage{amsmath}
\usepackage{esint}                   
\usepackage{epsfig}
\usepackage{latexsym}
\usepackage{color}
\usepackage{graphicx}
\usepackage{hyperref}
\usepackage{nicefrac}
\usepackage[latin1]{inputenc}
\usepackage{pstricks}
\usepackage{slashed}
\usepackage{multirow}
\usepackage{ulem}
\usepackage{comment}
\usepackage[nosort]{cite}
\usepackage{datetime}
\usepackage{tikz-cd}
\usepackage{float}

\def\hybrid{\topmargin -30pt    \oddsidemargin 0pt 
        \headheight 0pt \headsep 0pt
        \textwidth 6.25in       
        \textheight 9.5in       
        \marginparwidth .875in
        \parskip 5pt plus 1pt   \jot = 1.5ex}

\hybrid

\def\baselinestretch{1.2}

\catcode`\@=11

\def\marginnote#1{}
\newcount\hour
\newcount\minute
\newtoks\amorpm
\hour=\time\divide\hour by60
\minute=\time{\multiply\hour by60 \global\advance\minute by-\hour}
\edef\standardtime{{\ifnum\hour<12 \global\amorpm={am}%
        \else\global\amorpm={pm}\advance\hour by-12 \fi
        \ifnum\hour=0 \hour=12 \fi
        \number\hour:\ifnum\minute<10 0\fi\number\minute\the\amorpm}}
\edef\militarytime{\number\hour:\ifnum\minute<10 0\fi\number\minute}

\def\draftlabel#1{{\@bsphack\if@filesw {\let\thepage\relax
   \xdef\@gtempa{\write\@auxout{\string
      \newlabel{#1}{{\@currentlabel}{\thepage}}}}}\@gtempa
   \if@nobreak \ifvmode\nobreak\fi\fi\fi\@esphack}
        \gdef\@eqnlabel{#1}}
\def\@eqnlabel{}
\def\@vacuum{}
\def\draftmarginnote#1{\marginpar{\raggedright\scriptsize\tt#1}}

\def\draft{\oddsidemargin -.5truein
        \def\@oddfoot{\sl preliminary draft \hfil
        \rm\thepage\hfil\sl\today\quad\militarytime}
        \let\@evenfoot\@oddfoot \overfullrule 3pt
        \let\label=\draftlabel
        \let\marginnote=\draftmarginnote
   \def\@eqnnum{(\theequation)\rlap{\kern\marginparsep\tt\@eqnlabel}%
\global\let\@eqnlabel\@vacuum}  }

\def\draft2{
        \def\@oddfoot{\sl preliminary draft \hfil
        \rm\thepage\hfil\sl\today\quad\militarytime}
        \let\@evenfoot\@oddfoot \overfullrule 3pt
        \let\marginnote=\draftmarginnote
   \def\@eqnnum{(\theequation)\rlap{\kern\marginparsep\tt\@eqnlabel}%
\global\let\@eqnlabel\@vacuum}  }

\def\preprint{\twocolumn\sloppy\flushbottom\parindent 2em
        \leftmargini 2em\leftmarginv .5em\leftmarginvi .5em
        \oddsidemargin -.5in    \evensidemargin -.5in
        \columnsep .4in \footheight 0pt
        \textwidth 10.in        \topmargin  -.4in
        \headheight 12pt \topskip .4in
        \textheight 6.9in \footskip 0pt
        \def\@oddhead{\thepage\hfil\addtocounter{page}{1}\thepage}
        \let\@evenhead\@oddhead \def\@oddfoot{} \def\@evenfoot{} }

\def\numberbysection{\@addtoreset{equation}{section}
        \def\theequation{\thesection.\arabic{equation}}}

\def\underline#1{\relax\ifmmode\@@underline#1\else
        $\@@underline{\hbox{#1}}$\relax\fi}

\def\titlepage{\@restonecolfalse\if@twocolumn\@restonecoltrue\onecolumn
     \else \newpage \fi \thispagestyle{empty}\c@page\z@
        \def\thefootnote{\fnsymbol{footnote}} }

\def\endtitlepage{\if@restonecol\twocolumn \else \newpage \fi
        \def\thefootnote{\arabic{footnote}}
        \setcounter{footnote}{0}}  

\catcode`@=12
\relax

\def\figcap{\section*{Figure Captions\markboth
        {FIGURECAPTIONS}{FIGURECAPTIONS}}\list
        {Figure \arabic{enumi}:\hfill}{\settowidth\labelwidth{Figure
999:}
        \leftmargin\labelwidth
        \advance\leftmargin\labelsep\usecounter{enumi}}}
 \relax
\def\tablecap{\section*{Table Captions\markboth
        {TABLECAPTIONS}{TABLECAPTIONS}}\list
        {Table \arabic{enumi}:\hfill}{\settowidth\labelwidth{Table
999:}
        \leftmargin\labelwidth
        \advance\leftmargin\labelsep\usecounter{enumi}}}
 \relax
\def\reflist{\section*{References\markboth
        {REFLIST}{REFLIST}}\list
        {[\arabic{enumi}]\hfill}{\settowidth\labelwidth{[999]}
        \leftmargin\labelwidth
        \advance\leftmargin\labelsep\usecounter{enumi}}}
 \relax
\makeatletter
\newcounter{pubctr}
\def\publist{\@ifnextchar[{\@publist}{\@@publist}}
\def\@publist[#1]{\list
        {[\arabic{pubctr}]\hfill}{\settowidth\labelwidth{[999]}
        \leftmargin\labelwidth
        \advance\leftmargin\labelsep
        \@nmbrlisttrue\def\@listctr{pubctr}
        \setcounter{pubctr}{#1}\addtocounter{pubctr}{-1}}}
\def\@@publist{\list
        {[\arabic{pubctr}]\hfill}{\settowidth\labelwidth{[999]}
        \leftmargin\labelwidth
        \advance\leftmargin\labelsep
        \@nmbrlisttrue\def\@listctr{pubctr}}}
 \relax
\makeatother

\def\be{\begin{equation}}
\def\ee{\end{equation}}
\def\ba{\begin{eqnarray}}
\def\ea{\end{eqnarray}}

\def\del{\partial}

\def\k{\kappa}
\def\r{\rho}
\def\a{\alpha}

\def\b{\beta}

\def\g{\gamma}
\def\G{\Gamma}
\def\d{\delta}
\def\D{\Delta}
\def\e{\epsilon}

\def\th{\theta}

\def\m{\mu}
\def\n{\nu}
\def\om{\omega}
\def\Om{\Omega}
\def\l{\lambda}

\def\s{\sigma}

\def\cL{{\cal L}}

\def\no{\noindent}

\def\qq{\qquad}

\def\IR{\relax{\rm I\kern-.18em R}}

\def\inv{^{\raise.0ex\hbox{${\scriptscriptstyle -}$}\kern-.05em 1}}

\def \ov {\over}

\begin{document}


\renewcommand{\theequation}{\thesection.\arabic{equation}}
\csname @addtoreset\endcsname{equation}{section}

\begin{titlepage}
\begin{center}

\renewcommand*{\thefootnote}{\arabic{footnote}}

\phantom{xx}
\vskip 0.5in

{\large {\bf  Integrable models based on non-semi-simple groups \\ and
plane wave target spacetimes}}

\vskip 0.5in

{\bf  Konstantinos Sfetsos}\hskip .15cm and {\bf Konstantinos Siampos}

\vskip 0.11in

Department of Nuclear and Particle Physics, \\
Faculty of Physics, National and Kapodistrian University of Athens, \\
Athens 15784, Greece

\vskip .3 cm

{\footnotesize \texttt ksfetsos, konstantinos.siampos@phys.uoa.gr}


\vskip .2in

\end{center}

\vskip .2in

\centerline{\bf Abstract}
\no
 We initiate the construction of integrable $\lambda$-deformed WZW models based on non-semisimple groups. We focus on the four-dimensional case whose underlying symmetries are based on the non-semisimple group $E_2^c$. The corresponding gravitational backgrounds of Lorentzian signature are plane waves which can be obtained as Penrose limits of the $\lambda$-deformed $SU(2)$  background times a timelike coordinate for appropriate choices of the $\lambda$-matrix. We construct two such deformations which we demonstrate to be integrable. They both deform the Nappi--Witten plane wave and are inequivalent. Nevertheless,  they have the same underlying symmetry algebra which is a Saletan-type contraction of that for the $\lambda$-deformed $SU(2)$ background with a timelike direction. We also construct a plane wave from the Penrose limit of the $\lambda$-deformation of the  $\nicefrac{SU(2)}{U(1)}$ coset CFT times a timelike coordinate which represents the deformation of a logarithmic CFT constructed in the past. Finally, we briefly consider contractions based on the simplest Yang--baxter $\sigma$-models.

\vskip .4in

\vfill

\end{titlepage}
\vfill
\eject



\def\baselinestretch{1.2}
\baselineskip 20 pt

\newcommand{\eqn}[1]{(\ref{#1})}

\tableofcontents


\section{Introduction}

Two-dimensional non-linear $\s$-models describe propagating strings in curved spacetimes with action given by\footnote{The world-sheet coordinates $\s^\pm$ and the $(\tau,\s)$ are given in terms of
$$
\s^\pm=\tau\pm\s\,,\quad \del_\pm=\frac12\left(\del_\tau\pm\del_\s\right)\,,\quad d^2\s=d\tau\,d\s\,.
$$
\label{worldsheet.der}
}
\begin{equation}
\label{action}
S= \frac{1}{2\pi}\int_{\del B} d^2\sigma \cL\ ,\quad \cL = (G_{\m\n}+ B_{\m\n}) \del_+X^\m\del_-X^\n\, .
\end{equation}
From this one may read off the metric $G_{\mu\nu}$ as well as the antisymmetric tensor $B_{\mu\nu}$.  A large class of consistent (conformal) backgrounds can be constructed using WZW models which are based on simple or semi-simple groups~\cite{Witten:1983ar}. In constructing the corresponding actions a key ingredient is the invertibility of the Killing form $\kappa_{AB}\propto f_{AC}{}^D f_{BD}{}^C$. For non-semi-simple group the Killing form $\kappa_{AB}$ is degenerate (non-invertible) and an alternative definition is required. Nappi and Witten showed us in~\cite{Nappi:1993ie} how to address this problem for the example of the centrally extended two-dimensional Euclidean group $E_2^c$. The derived background assumes the form of a plane wave and can be obtained as a Penrose-type of the $SU(2)$ WZW model times a timelike coordinate~\cite{Sfetsos:1993rh} or equivalently as a Saletan-type contraction ~\cite{Saletan:1961} on the underlying current symmetry algebra~\cite{Olive:1993hk}.\footnote{Some selected past works on the subject elucidating various aspects can be found in~\cite{Kiritsis:1993jk,Sfetsos:1993na,Kehagias:1994iy,Antoniadis:1994jx,Sfetsos:1994fc,Figueroa-OFarrill:1995lva,Bianchi:2004vf}.}

\no
In this article we will extend the work of~\cite{Nappi:1993ie} to current bilinear, at the infinitesimal level, deformations of the WZW model, which generically are known as $\l$-deformed models~\cite{Sfetsos:2013wia}. These models interpolate from the WZW in the UV to the non-Abelian T-dual of the PCM towards the IR~\cite{Sfetsos:2013wia,Itsios:2014lca,Sfetsos:2014jfa,Appadu:2015nfa} and for specific choices of
 the deformation matrix $\l_{ab}$ they preserve integrability~\cite{Sfetsos:2013wia,Balog:1993es,Hollowood:2014rla,Itsios:2014vfa,Sfetsos:2015nya,Sfetsos:2014lla}. With the present work we will bring into this general framework the notion of Penrose limit and the corresponding Saletan-type contractions. We will derive backgrounds with the spacetime interpretation of plane waves that can be obtained as Penrose-type limits on $\l$-deformed backgrounds, thus extending the work of~\cite{Sfetsos:1993rh} for the WZW case. In addition, the underlying symmetry algebras will be  derived as a Saletan-type contraction~\cite{Saletan:1961} on the underlying symmetry algebra of the $\l$-deformed models~\cite{Balog:1993es,Sfetsos:2013wia,Georgiou:2016iom}.

\no
This work is structured as follows: In Section~\ref{Sec2}, we review the WZW model on the non-semi-simple group $E_2^c$, which is the central extension of the Euclidean group $E_2$. Then, we consider its $\l$-deformations for isotropic and non-invertible deformation matrices, leading to distinct background solutions. In both cases connection with plane wave limits of the isotropic $\l$-deformed $SU(2)$ times a timelike scalar, performed in \cite{Georgiou:2022fow}, is made. 
In Section~\ref{Sec3}, we prove that the  isotropic and the non-invertible deformations are integrable as their equations take the form of a flat Lax connection. In addition, we show that the corresponding conserved charges are in involution as the Poisson brackets of the spatial Lax component assume the Maillet form~\cite{Maillet:1985ek,Maillet:1985ec}. In addition, we derive the symmetry algebras of the $\l$-dressed perturbing operators. In Section~\ref{Section.algebra}, we consider the Saletan-type  contraction on the symmetry algebras of the $\l$-deformed $SU(2)$ model times a timelike direction matching the symmetry algebra of the corresponding $\l$-deformed $E_2^c$ model. Next, we consider the Saletan type construction on the symmetry algebra induced by a Penrose limit and show that they are in one to one correspondence and compatible.
Finally, in Appendix~\ref{AppA} we consider various Penrose limits on $\l$- and Yang--Baxter-deformed backgrounds times a timelike direction. 
A particularly notable case corresponds to the deformation of a logarithmic CFT constructed  in \cite{Bakas:2002qh}, via a Penrose limit of the $\nicefrac{SU(2)}{U(1)}$ coset CFT times a timelike coordinate.

\section{$\l$-deformations based on non-semi-simple groups}
\label{Sec2}

\no 
In this Section we first review in our context the WZW model for the non-semi-simple group $E^c_2$~\cite{Nappi:1993ie}. Then we consider their $\l$-deformations for isotropic and non-invertible matrices. The result will be two inequivalent plane wave spacetimes.

\subsection{WZW model on $E^c_2$}
\label{WZWE2C}

The action of the WZW model for a group $G$ and the level $k$ is given by
\be
\label{WZW.action.singular}
S_k(g)=\frac{k}{2\pi}\int_{\del B} d^2\s\, \Om_{AB} L^A_+ L^B_--\frac{k}{24\pi}\int_B f_{AB}{}^D\Om_{DC}\,R^A\wedge R^B\wedge R^C\,,
\ee
where the right- and the left-invariant Maurer--Cartan one-forms $R^A, L^A$ as well as the adjoint action $D^A{}_B$ are defined in terms of 
\be
\label{RL.MC.forms}
\begin{split}
&R=R^AT_A=-idgg^{-1}\,,\quad L=L^AT_A=-ig^{-1}dg\,,\\
&R_{A\pm}=-i\, \text{Tr}(T_A \del_\pm g g^{-1})\,,\quad L_{A\pm}=-i\, \text{Tr}(T_A g^{-1}\del_\pm g )\,,\\
& D^A{}_BT_A=gT_Bg^{-1}\,,\quad [T_A,T_B]=if_{AB}{}^CT_C\,,\quad \text{Tr}(T_AT_B)=\Om_{AB}
\end{split}
\ee
and $T_A$'s are the Hermitian generators of the group $G$. Let us now focus on the non-semi-simple group $E^c_2$, whose algebra of generators 
$T_A=\{iP_1,iP_2,iJ,iF\}$, with $A=1,2,3,4$, is given in terms of
\be
\label{Ec2.commutators}
[J, P_i]=\e_{ij}P_j\ , \qq [P_i, P_j]=\e_{ij}F \ ,\qq [F,J]=[F,P_i]=0\,,
\ee
yielding the structure constants
\be
\label{structure.constants}
f_{12}{}^4=1\,,\quad f_{13}{}^2=-1\,,\quad f_{23}{}^1=1\,.
\ee
Unlike the case of semi-simple algebras, here the Killing form $\kappa_{AB}\propto f_{AC}{}^D f_{BD}{}^C$ is degenerate. Hence, following~\cite{Nappi:1993ie} we introduce 
a quadratic form $\Om_{AB}$ satisfying
\be
\label{Omega.properties}
\Om_{AB}=\Om_{BA}\,,\quad f_{AB}{}^D\Omega_{DC}+ f_{AC}{}^D\Omega_{BD} = 0\,,\quad  \Om_{AC}\Om^{CB}=\d_A{}^B\,,
\ee
where the first and the second property in \eqref{Omega.properties} ensure the existence of the quadratic and the Wess--Zumino term in the WZW action~\eqref{WZW.action.singular}, whereas the third one provides a way to properly lower and raise indices. For the case at hand using \eqref{Ec2.commutators} and \eqref{Omega.properties} we find~\cite{Nappi:1993ie}
\be
\label{Omega.Inv}
\Omega_{AB} =
\begin{pmatrix}
1 & 0 & 0 & 0\\ 
0 & 1 & 0 & 0\\ 
0 & 0 & b & 1\\ 
0 & 0 & 1 & 0
\end{pmatrix}\,,\quad
\Omega^{AB} =
\begin{pmatrix}
1 & 0 & 0 & 0\\ 
0 & 1 & 0 & 0\\ 
0 & 0 & 0 & 1\\ 
0 & 0 & 1 & -b
\end{pmatrix}\,,
\ee
where $b$ is an arbitrary constant. We also parametrize the group element as 
\be
\label{group.uv}
g= e^{ia_i P_i}e^{2iuJ+\frac{i\,v}{2}F}\,,
\ee
yielding the left-, right-invariant Maurer--Cartan one-forms derived from \eqref{RL.MC.forms}\footnote{Making use of the Baker--Campbell--Hausdorff identity
$$
\del e^A=\int_0^1dx\, e^{xA}\,\del A\, e^{(1-x)A}\,,\quad e^XY\,e^{-X}=Y+[X,Y]+\frac{1}{2}[X,[X,Y]]+\frac{1}{6}[X,[X,[X,Y]]]+\cdots
$$}
\be
\label{LR.E2C}
\begin{split}
&L=(d a_i\cos 2u+\epsilon_{ij}d a_j\sin 2u)P_i+\frac12(d v-\epsilon_{ij}a_id a_j)F+2d u J \ ,\\ 
&R= (d a_i+2\epsilon_{ij}a_jd u)P_i+\frac12\left(d v+\epsilon_{ij}a_id a_j-2a_ia_idu\right)F+2d u J \,.
\end{split}
\ee
In addition, we compute the adjoint action \eqref{RL.MC.forms}
\be
\label{D.E2C}
D^A{}_B=\begin{pmatrix}
\cos 2u & -\sin 2u & a_2 & 0\\ 
\sin 2u & \cos 2u & -a_1 & 0\\ 
0 & 0 & 1 & 0\\ 
a_1\sin 2u-a_2 \cos 2u & a_2\sin 2u+a_1 \cos 2u & -\frac12a_ia_i & 1
\end{pmatrix},
\ee
satisfying the identities 
\be
\label{D.identities}
\begin{split}
&\Om_{CD}D^C{}_AD^D{}_B=\Om_{AB}\,,\quad D^A{}_C\Om^{CD}D^B{}_D=\Om^{AB}\,,\quad
R^A=D^A{}_B L^B\,,\\
&f_{AB}{}^D\Om_{DC}=D^E{}_AD^F{}_BD^G{}_Cf_{EF}{}^H\Om_{HG}\,,
\end{split}
\ee
where we denote $\Om^{AB}=(\Om^{-1})_{AB}$.
Then, the WZW Lagrangian \eqref{WZW.action.singular} in the normalization of \eqref{action} takes the form\footnote{In the normalization of \eqref{action} we can read the line element and the field strength of the antisymmetric field $H=dB$, matching Eq. (17) and $H_{ua_1a_2}=1$ in~\cite{Nappi:1993ie}, up to the overall level $k$ and the identification $(a_1,a_2,u,v)_\text{here}\to(a_1,a_2,\nicefrac{u}{2},2v)_\text{there}$.}
\be
\label{Lag.density.WZW}
{\cal  L}_k=k \big( \del_+ a_i \del_- a_i
+ \del_+ u \del_- v + \del_+ v \del_- u + 4b\ \del_+ u \del_- u
+2 \e_{ij}\del_+ u \del_- a_i a_j  \big).
\ee

\no
Then, changing variables as
\be
\label{atox0}
a_1= x_2 + x_1 \cos 2u, \quad a_2= x_1 \sin 2u,
\quad v\to v + x_1 x_2 \sin 2u
\ee
and adding the total derivative term
\be
\label{derivative0}
\frac{k}{4\pi}d\left(-\frac12\cos 2u(x_1dx_2-x_2dx_1)+udv\right),
\ee
the WZW Lagrangian density \eqref{Lag.density.WZW} is expressed in terms of Rosen coordinates~\cite{Nappi:1993ie}\footnote{In the normalization of \eqref{action}, the corresponding line element matches Eq. (26) in~\cite{Nappi:1993ie}, up to the overall level $k$ and the identification $(x_1,x_2,u,v)_\text{here}\to(f,a,\nicefrac{u}{2},2v)_\text{there}$.}
\be
\label{CFT.E2C}
{\cal L}_k=k \big( \del_+ x_1 \del_- x_1 +\del_+ x_2 \del_- x_2 
+ 2 \cos 2u\ \del_+ x_1 \del_- x_2 + 4b\ \del_+ u \del_- u +2 \del_+ v \del_- u \big),
\ee
where the constant $b$ can be set to zero if we shift $v\to v-2bu$.

\subsection{Deforming with an isotropic $\l$-matrix}
\label{isotropic.section}

We next extend the construction of~\cite{Nappi:1993ie} summarized in Section~\ref{WZWE2C} for the $\l$-deformed models of~\cite{Sfetsos:2013wia} 
\be
\label{lambda.action.singular}
S_\l(g)=S_k(g)+\frac{k}{\pi}\int_{\del B} d^2\s\, R^A_+[(\mathbb{I}-\l D^T)^{-1}\l]_A{}^C\Om_{CB} L^B_-\,,
\ee
where $S_k(g)$ is the WZW action \eqref{WZW.action.singular}, for the non-semi-simple group $E_2^c$.

\no
Choosing as deformation matrix 
$
 \lambda_A{}^B=\l\d_A{}^B
$,
and the results of Section~\ref{WZWE2C}, we may compute the  Lagrangian density for \eqref{lambda.action.singular} (in the normalization of \eqref{action}) to be
\be
\label{lambda.non-semisimple}
\begin{split}
{\cal L}_\l&=k\left\{\frac{1-\l^2}{\D(u)}\del_+ a_i\del_- a_i+\frac{2\l\sin 2u}{\D(u)}\epsilon_{ij}\del_+ a_i\del_- a_j\right.\\
&+\frac{1+\l}{1-\l}( \del_+ u \del_- v + \del_+ v \del_- u )+4\frac{1+\l}{1-\l}\left(b-\l\frac{a_ia_i}{\D(u)}\right)\del_+u\del_-u\\
&+2\epsilon_{ij}\frac{1+\l}{\D(u)}(a_j\del_-a_i\del_+u-\l a_j\del_+a_i\del_-u)\\
&\left.+\frac{4\l\sin 2u}{(1-\l)\D(u)}(a_i\del_- a_i\del_+ u+\l a_i\del_+ a_i\del_- u)\right\}\,,
\end{split}
\ee
where 
\be
\label{Du}
\D(u)=1+\l^2-2\l\cos 2u\ .
\ee 
As in the $\l=0$ case the constant $b$ can be absorbed by shifting $v\to v-2bu$. From the above Lagrangian density and \eqref{action} we can easily read the metric and the antisymmetric tensor. The resulting $\s$-model can be made, if one wishes, conformal  by adding the dilaton (at one-loop order)
\be
\label{isotr.dilaton}
\Phi(u)=-\frac12\ln\D(u)+\frac{2\l^2u^2}{(1-\l^2)^2}\,.
\ee
It can be checked that the corresponding metric read from \eqref{lambda.non-semisimple} using \eqref{action},  describes a conformally flat plane wave background, where $\xi=\del_v$ is a null-geodesic vector and, as all plane-waves, has vanishing scalar invariants. 

\no
Let us now express \eqref{lambda.non-semisimple} in Brinkmann coordinates which can be done in three steps. 
In the first step we change coordinates, generalizing \eqref{atox0} in the presence of $\l$ 
\be
\label{atoxl}
\begin{split}
&a_1= x_2 + x_1 \cos 2u, \quad a_2= x_1 \sin 2u,\\
&v\to v +x_1 x_2\sin 2u f_1(u)+(x_1^2+x_2^2)f_2(u)\,,\\
&f_1(u)=\frac{(1-\l)^2-2\l\cos 2u}{\D(u)}\,,\quad f_2(u)=-\frac{\l\sin 2u}{\D(u)}\, .
\end{split} 
\ee
We also extend \eqref{derivative0} by adding the total derivative term
\be
\label{derivativel}
\frac{k}{4\pi}d\left(-\frac12 f_3(u)(x_1dx_2-x_2dx_1)+\frac{1+\l}{1-\l}udv\right)\,,\quad f_3(u)=\cos 2u+2\ln\D(u)\,.
\ee
The second step is to diagonalize the corresponding metric in the $(x_1,x_2)$ plane 
\be
\label{rotate12}
x_1\to\frac{1}{\sqrt{2}}(x_1-x_2)\,,\quad x_2\to\frac{1}{\sqrt{2}}(x_1+x_2)\ .
\ee 
Finally, a last coordinate change 
\be
\begin{split}
&x_1\to\frac{x_1}{\sqrt{f_4(u)}}\,,\quad x_2\to\frac{x_2}{\sqrt{f_5(u)}}\,,
\\ 
&v\to v+\frac14\left(\frac{f_4'(u)}{f_4(u)}x_1^2+\frac{f_5'(u)}{f_5(u)}x_2^2\right)+f_6(x_1,x_2,u)\, ,
\\
&f_4(u)=\frac{2(1-\l^2)\cos^2u}{\D(u)}\,,\quad f_5(u)=\frac{2(1-\l^2)\sin^2u}{\D(u)}\,,
\\
&f_6(x_1,x_2,u)=\frac{\l}{(1+\l)\D(u)}\left((1+\l)^2x_1^2\tan u-(1-\l)^2x_2^2\cot u\right)\,,
\end{split}
\ee
brings the metric (read from \eqref{action}) into the Brinkmann form
\be
\label{gpp.su21}
\begin{split}
&ds^2=2\frac{1+\l}{1-\l}dudv+dx_1^2+dx_2^2+\left(4b\frac{1+\l}{1-\l}+(x_1^2+x_2^2)F(u)\right)du^2\,,\\
&F(u)=- {(1-\l)^2\ov \D^2(u)}\Big((1+10\l + \l^2)\cos^2u+ (1+\l)^2\sin^2u\Big)\ .
\end{split}
\ee
The field strength of the antisymmetric tensor reads
\be
\label{Bpp.su21}
H= dB= -2\left( {2\l\ov 1-\l^2} + {1-\l^2\ov\D(u)}\right) du\wedge dx_1\wedge dx_2\,.
\ee
The parameter $b$ in \eqref{gpp.su21} can be set to zero similar to what was noted below \eqn{Du}.
The above background is invariant under the non-perturbative symmetry
\be
\label{syymm}
\l\to\frac{1}{\l}\,,\qq u\to - u\ ,
\ee
This symmetry is in accordance with the similar symmetry  of the $\l$-deformed $\s$-model \eqn{lambda.action.singular} based on a (non-)semi-simple group $G$ \cite{Itsios:2014lca}.

\no
It can be checked that the above background can be obtained by a Penrose limit as in \cite{Georgiou:2022fow} which we review for completeness. First we consider the $\l$-deformed $SU(2)$ case, whose metric and the antisymmetric tensor are given by~\cite{Balog:1993es,Sfetsos:2013wia}
\be
\label{g.su2}
\begin{split}
&ds^2=2k\left(\frac{1+\l}{1-\l} d\a^2+\frac{1-\l^2}{\D(\a)}\sin^2\a\Big(d\beta^2+\sin^2\beta\, d\gamma^2\Big)\right)\,,\\
&\D(\a)=(1+\l)^2\sin^2\a+(1-\l)^2\cos^2\a\,
\end{split}
\ee
and 
\be
\label{B.su2}
B=2k\left(-\a+\frac{(1-\l)^2}{\D(\a)}\sin\a\cos\a\right)\sin\beta\, d\beta\wedge d\gamma\,,
\ee
where the range of the parameter $\l$ is $0\leqslant \l <1$. 

\no
To proceed, we add to \eqn{g.su2} the term $-2 k\, dt^2$ and we let
\be
\label{pensorel2}
\a=u\, , \quad t=  \sqrt{1+\l\ov 1-\l}\, u -{1\ov 2k}  \sqrt{1-\l\ov 1+\l}\, v\,, \quad \b ={\r\ov \sqrt{2k}}\,.
\ee
Then, in the $k\to \infty$ limit we obtain a plane wave line element in Rosen coordinates 
\be
\label{gpp.su2}
\begin{split}
&ds^2=2 du dv + {1-\l^2\ov \D(u)}\sin^2u (dx_1^2 + dx_2^2) \ ,
\\
& \D(u)=(1-\l)^2\cos^2u+(1+\l)^2\sin^2u\,
\end{split}
\ee
and the antisymmetric tensor
\be
\label{Bpp.su2}
B=\Big(\!-u+ {(1-\l)^2\ov  \D(u)}\sin u\cos u\Big)\ dx_1 \wedge dx_2 \, ,
\ee
where we have passed from the polar coordinates $(\r,\g)$ to Cartesian ones $(x_1,x_2)$.

\no
Transforming into Brinkmann coordinates 
\be
x_i\to\frac{x_i}{\sqrt{f(u)}}\,,\quad  v\to v+\frac{f'(u)}{4f(u)}(x_1^2+x_2^2)\,,\quad f(u)=\frac{1-\l^2}{\D(u)}\sin^2u\,,
\ee
then \eqref{gpp.su2} and \eqref{Bpp.su2} becomes precisely \eqn{gpp.su21} (with $b=0$) and \eqn{Bpp.su21}, respectively.

\subsection{Deforming with an non-invertible $\l$-matrix}
\label{constrained.section}

The tentative reader might have noticed that the $\l$-deformed action \eqn{lambda.action.singular} was written in such a way that one may consider models in which the matrix $\l$ is non-invertible. In that respect, we may consider the matrix
\be
\label{matrix.lambda.non}
\l_A{}^B=\text{diag}(\l,-\l,0,0)\, ,
\ee 
hence the  deformation at linear order in $\l$ is driven by the generators $P_1$ and $P_2$. These, according to \eqn{Ec2.commutators} do not form an Abelian sub-algebra but since their commutator is the central element $F$, thus commuting with all generators, the perturbation is a consistent one beyond the linear order. The relative coefficient is minus so that the resulting background is as simple as possible for deformations in which the matrix $\l$ is non-invertible.

\no
In this case and for comparison with the recent literature~\cite{Georgiou:2022fow}, we identify $u=\nicefrac{x^+}{2},v=2x^-$. Then using the results of Section~\ref{WZWE2C} into \eqref{lambda.action.singular} yields the Lagrangian density (in the normalization of \eqref{action})
\be
\label{lambda.action.singular1}
\begin{split}
&{\cal L}_\l=
k\left\{\frac{\D_+(x^+)}{1-\l^2}\del_+a_1\del_-a_1+\frac{\D_-(x^+)}{1-\l^2}\del_+a_2\del_-a_2\right.\\
&+\frac{2\l\sin u}{1-\l^2}\left(\del_+a_1\del_-a_2+\del_+a_2\del_-a_1\right)+\del_+x^+\del_-x^-+\del_+x^-\del_-x^++b\del_+x^+\del_-x^+\\
&\left.+\frac{a_2\D_+(x^+)-2a_1\l\sin x^+}{1-\l^2}\del_+x^+\del_-a_1-\frac{a_1\D_-(x^+)-2a_2\l\sin x^+}{1-\l^2}\del_+x^+\del_-a_2\right\}\,,
\end{split}
\ee
where $\D_\pm(x^+)=1+\l^2\pm2\l\cos x^+$.\footnote{As in the isotropic case the resulting $\s$-model can be made conformal with the addition of the appropriate dilaton.} As in the case of the isotropic deformation, the constant $b$ can be set to zero by shifting $x^-\to x^--\frac{b}{2}x^+$. 
It can be  checked that the metric corresponding to \eqn{lambda.action.singular1}  describes a plane wave background, $\xi=\del_v$ is a null-geodesic vector and has vanishing scalar invariants. Unlike the isotropic case, the background is non-conformally flat (non-vanishing Weyl tensor of the Petrov type $N$) for $\l\neq0$. Hence, the isotropic and the non-invertible $\l$ deformations of $E_2^c$, given in terms of the actions \eqref{lambda.non-semisimple} and \eqref{lambda.action.singular1} respectively, are not equivalent, under diffeomorphisms, when $\l\neq0$. In fact they are not equivalent, as we will see in Section~\ref{Sec:Symmetry} under canonical transformations in phase space as well. 

\no
Let us now express the corresponding metric into the Brinkmann form. As before this can be done in three steps. In the first step we extend \eqref{atox0} in the presence of $\l$ as 
\be
\begin{split}
&a_1=x_2+x_1\cos x^+\,,\quad a_2=x_1\sin x^+\,,\\ 
&x^-\to x^-+\frac12\times\frac{1+\l^2}{1-\l^2}\left(x_1x_2+\frac{\l}{1+\l^2}(x_1^2+x_2^2)\right)\sin x^+\, .
\end{split}
\ee
We also extend \eqref{derivative0} by adding the total derivative term
\be
\frac{k}{4\pi}d\left(-\frac{1+\l^2}{2(1-\l^2)}\cos x^+(x_1dx_2-x_2dx_1)+x^+dx^-\right).
\ee
The second step is to bring the corresponding metric in the diagonal form in the $(x_1,x_2)$ plane through a rotation
\be
x_1\to\frac{1}{\sqrt{2}}(x_1-x_2)\,,\quad x_2\to\frac{1}{\sqrt{2}}(x_1+x_2)\ .
\ee 
Finally, the last coordinate change 
\be
\begin{split}
&x_1\to\frac{x_1}{\sqrt{f_+(x^+)}}\,,\quad x_2\to\frac{x_2}{\sqrt{f_-(x^+)}}\,,\quad f_\pm(x^+)=\frac{1\pm\l}{1\mp\l}(1\pm\cos x^+)\ ,
\\
&v\to v+\frac14\left(\frac{f'_+(x^+)}{f_+(x^+)}x_1^2+\frac{f'_-(x^+)}{f_-(x^+)}x_2^2\right)\,,
\end{split}
\ee
brings the line element into the Brinkmann form (read from \eqref{action})
\be
\label{metric.singular}
ds^2=2dx^+dx^-+dx_1^2+dx_2^2-\frac14\left(\Big(\frac{1-\l}{1+\l}\Big)^2x_1^2
+\Big(\frac{1+\l}{1-\l}\Big)^2x_2^2\right)(dx^+)^2\ ,
\ee 
where we have shifted $b$ to zero and the field strength of the antisymmetric tensor to the form
\be
\label{H.singular}
H=-\frac{1+\l^2}{1-\l^2}\, dx^+\wedge dx_1\wedge dx_2\,.
\ee
Similarly to the plane wave we examined previously, this one is invariant under the non-perturbative symmetry
\be
\label{syymm1}
\l\to\frac{1}{\l}\,,\qq x^\pm \to -  x^\pm\ ,
\ee
It can be proved that the background \eqref{metric.singular} can be derived using a Penrose limit~\cite{Georgiou:2022fow}.
In particular, adding to \eqref{g.su2}, the term $\displaystyle - 2 k {1 - \l\ov 1 + \l} dt^2$, the resulting four-dimensional spacetime has then an obvious null geodesic given by
\be 
\label{geo}
\a=\b={\pi\ov 2}\ ,\quad t=\g\ .
\ee
Let us now consider the following zoom-in limit~\cite{Georgiou:2022fow}\footnote{This zoom-in limit was first given in Eq.(2.7) of~\cite{Georgiou:2022fow}, upon  rescaling $\displaystyle x^\pm\to \sqrt{1\pm\l\ov 1\mp\l}x^\pm$.}
\be
\label{pensorel}
\begin{split}
& t=\frac12\left(\frac{1+\l}{1-\l}x^+-\frac1k x^-\right)\ ,\quad \g =\frac12\left(\frac{1+\l}{1-\l}x^++\frac1k x^-\right)\ ,
\\
&
\a = {\pi\ov 2} + \sqrt{1-\l\ov 1+\l}\, \frac{x_1}{\sqrt{2 k}}\ ,\qq  \b = {\pi\ov 2} + \sqrt{1+\l\ov 1-\l}\, \frac{x_2}{\sqrt{2k}}\ .
\end{split}
\ee
In the limit of $k\to \infty$, the metric and the field strength of the antisymmetric tensor are given in terms of 
\eqref{metric.singular} and \eqref{H.singular} respectively.

\section{Integrability and symmetry algebras}
\label{Sec3}

In this Section we prove that the deformations we have constructed are integrable by showing that their equation of motion assume the form of a flat Lax connection. Then, we show that the conserved charges are in involution as the Poisson brackets of the spatial Lax component assume the Maillet form~\cite{Maillet:1985ek,Maillet:1985ec}. Finally, we work out the corresponding symmetry algebras for the aforementioned deformations. 

\subsection{The Lax connection}

Our starting point are the equations of motion for the Lagrangian density \eqref{lambda.action.singular} which take the compact form that is adequate also for non-semi-simple 
groups~\cite{Sfetsos:2013wia,Hollowood:2014rla}\footnote{In components the various insertions read
\begin{equation*}
\begin{split}
&A_\pm=A^A_\pm\,T_A\,,\quad \l^TA_+=\l^A{}_BA_+^BT_A\,,\quad \Om^{-1}\l\Om A_-=\Om^{AB}\l_B{}^C\Om_{CD}A_-^DT_A\,,\\
& [\l^TA_+,A_-]=f_{AB}{}^C(\l^TA_+)^AA_-^BT_C\,,\quad [A_+,\Om^{-1}\l\Om A_-]=f_{AB}{}^CA_+^A(\Om^{-1}\l\Om A_-)^BT_C\,,
\end{split}
\end{equation*}
where $T_A=\{iP_1,iP_2,iJ,iF\}$ are the generators $E_2^c$.}
\be
\begin{split}
&\del_+(\Om^{-1}\l\Om A_-)-\del_-A_+=[A_+,\Om^{-1}\l\Om A_-]\,,\\
&\del_+A_--\del_-(\l^TA_+)=[\l^TA_+,A_-]\,,
\end{split}
\ee
where
\be
\label{As.singular}
A_+=(\mathbb{I}-D\l^T)^{-1}\del_+gg^{-1}\,,\quad A_-=-\Om^{-1}(\mathbb{I}-D^T\l)^{-1}\Om g^{-1}\del_-g\,.
\ee
\paragraph{Isotropic $\l$-matrix:} Specializing to the $E_2^c$ group \eqref{Ec2.commutators} 
the structure constants were given in \eqref{structure.constants} and we can raise/lower indices using \eqref{Omega.Inv}. For the deformation matrix $\l_A{}^B=\l\d_A{}^B$, considered in Section~\ref{isotropic.section}, we find the equations of motion in terms of $A_\pm$
\be
\label{eom.isotropic}
\del_\pm A_\mp=\pm\frac{\l}{1+\l}[A_+,A_-]\,,
\ee
describing a classically integrable system since it admits a flat Lax connection~\cite{Sfetsos:2013wia, Balog:1993es,Hollowood:2014rla,Itsios:2014vfa}\footnote{\label{general.iso}In fact the Lax connection appearing in \eqref{ooskqsps} is valid for a general semi-simple or non-semisimple group, equipped with an $\Omega_{AB}$ which obeys \eqref{Omega.properties}.}
\be
\label{ooskqsps}
\del_+{\cal L}_--\del_-{\cal L}_+=[{\cal L}_+,{\cal L}_-]\,,\quad
{\cal L}_\pm=\frac{2\l}{1+\l}\frac{z}{z\mp1}A_\pm\,,
\ee
where $z$ is a complex spectral parameter. 
\paragraph{Non-invertible $\l$-matrix:} Using \eqref{structure.constants} for the deformation matrix $\l_A{}^B=\text{diag}(\l,-\l,0,0)$, considered in Section~\ref{constrained.section}, we find the equations of motion in terms of $A_\pm$,\footnote{In fact we could also consider as a flat Lax connection 
\begin{equation*}
{\cal L}_\pm=\sqrt{2\l}z^{\pm1}(A_\pm^1T_1\pm A^2_\pm T_2)\pm\l\frac{z^{\pm2}-\l}{1-\l^2}A_\pm^3T_3\pm(1+\zeta) A^4_\pm T_4\,,
\end{equation*}
where $\zeta$  may be thought of as an additional complex spectral parameter. This arbitrariness is due to the fact that $T_4=iF$ is the central element of the algebra $E_2^c$ commutes with all generators.}
\label{footLax}
\ba
\label{eom.As}
&&\del_\pm A^1_\mp=-\frac{\l}{1-\l^2}\left(A_\pm^2A_\mp^3+\l A_\mp^2A_\pm^3\right)\,,\quad
\del_\pm A^2_\mp=-\frac{\l}{1-\l^2}\left(A_\pm^1A_\mp^3-\l A_\mp^1A_\pm^3\right)\,,\nonumber\\
&&\del_\pm A^3_\mp=0\,,\quad \del_\pm A_\mp^4=\l(A_+^1A_-^2+A_-^1A_+^2)\, .
\ea
This leads to a classically integrable system since it admits the flat Lax connection
\be
\label{LAX.singular}
{\cal L}_\pm=\sqrt{2\l}z^{\pm1}(A_\pm^1T_1\pm A^2_\pm T_2)\pm\l\frac{z^{\pm2}-\l}{1-\l^2}A_\pm^3T_3\pm A^4_\pm T_4\,,
\ee
where $z$ is a complex spectral parameter and we recall the definition of $T_A=\{iP_1,iP_2,iJ,iF\},$ obeying the commutation relations \eqref{Ec2.commutators}.

\no
As a consistency check we have explicitly verified that the equations of motion \eqref{eom.isotropic} and \eqref{eom.As} (with the help of \eqref{LR.E2C}, \eqref{D.E2C}) are satisfied for the corresponding backgrounds \eqref{lambda.non-semisimple} and \eqref{lambda.action.singular1} respectively
\be
\del_+\del_-X^\mu+\left(\G_{\k\l}{}^\mu-\frac12H_{\k\l}{}^\mu\right)\del_+X^\k\del_-X^\l=0\,,\quad X^\mu=(a_1,a_2,u,v)\,,
\ee
where $\G_{\k\l}{}^\mu$ are the Christoffel symbols of $G_{\mu\nu}$ and $H_{\k\l\m}=\del_\k B_{\l\m}+\del_\m B_{\k\l}+\del_\l B_{\m\k}$, is the field strength of $B_{\mu\nu}$.

\subsection{ Hamiltonian integrability}
\label{Sec:Hamiltonian}

Given a flat Lax connection we can define the conserved monodromy matrix~\cite{Zakharov.Mikhailov}
\be
\del_\tau M(z)=0\,,\quad
M(z)=\text{Pexp}{\int_{-\infty}^\infty d\s{\cal L}_1}\,,\quad {\cal L}_1={\cal L}_+-{\cal L}_-\,,\quad \forall\, z\in\mathbb{C}\,.
\ee
Expanding the monodromy matrix in powers of $z$ introduces infinite conserved charges which are not a priori in involution. As it was proven in~\cite{Maillet:1985ek,Maillet:1985ec}, the conserved charges are in involution provided that
\be
\label{PB.spatial}
\{{\cal L}_1^{(1)}(\sigma_1;z),{\cal L}_1^{(2)}(\sigma_2;w)\}_\text{P.B.}=\left([r_{-zw},{\cal L}_1^{(1)}(\s_1;z)]+[r_{+zw},{\cal L}_1^{(2)}(\s_2;w)]\right)\d_{12}-2s_{zw}\d'_{12}\,,
\ee
where the superscript in parenthesis refer to the vector spaces on which the matrices act.\footnote{
In particular
\begin{equation*}
\begin{split}
&M^{(1)}=M^A T_A\otimes\mathbb{I}\,,\quad M^{(2)}=M^A\mathbb{I}\otimes T_A\\
&m^{(12)}=m^{AB} T_A\otimes T_B\otimes\mathbb{I}\,,\quad 
m^{(23)}=m^{AB} \mathbb{I}\otimes T_A\otimes T_B\,,\quad 
m^{(13)}=m^{AB} T_A\otimes\mathbb{I}\otimes T_B\,,
\end{split}
\end{equation*}
for an arbitrary vector $M=M^AT_A$ and a tensor $m=m^{AB}T_A\otimes T_B$.
}
In addition, $\d_{12}=\d(\s_1-\s_2) $ and $ \d'_{12}=\del_{\s_1}\d(\s_1-\s_2)$ correspond to ultralocal and non-ultralocal terms, respectively. The $r_{\pm zw}=r_{zw}\pm s_{zw}$, are matrices on the basis $T_A\otimes T_B$ with components $r_{\pm zw}^{AB}$, depending on $z$ and $w$. As a consequence of the anti-symmetry of the Poisson brackets \eqref{PB.spatial} they obey $r^{AB}_{+zw}+r^{BA}_{-wz}=0$ and consistency with the Jacobi identity yields the modified Yang--Baxter (mYB) relation
\be
\label{mYB.commutators}
[r_{+z_1z_3}^{(13)},r_{-z_1z_2}^{(12)}]+[r_{+z_2z_3}^{(23)},r_{+z_1z_2}^{(12)}]+[r_{+z_2z_3}^{(23)},r_{+z_1z_3}^{(13)}]=0\,.
\ee

\no
For the Lax pairs under consideration \eqref{ooskqsps} and  \eqref{LAX.singular}, to evaluate the Poisson brackets of their spatial Lax component in \eqref{PB.spatial}, we express the gauge fields $A_\pm$ in terms of the ${\cal S}_\pm$'s  satisfying the two commuting algebras~\cite{Bowcock:1988xr,Hollowood:2014rla}
\be
\label{djskwjdk0}
\{{\cal S}^A_\pm,{\cal S}^B_\pm\}_\text{P.B.}=\frac{2}{k}\left(f^{AB}{}_C\, {\cal S}^C_\pm\d_{12}\pm \Om^{AB}\d'_{12}\right)\,,
\ee
where
\be
\label{djskwjdk}
{\cal S}^A_+=A^A_+-(\Om^{-1}\l\Om)^A{}_B A^B_-\,,\quad {\cal S}^A_-=A^A_--(\l^T)^A{}_B A^B_+\,,
\ee
while the structure constants of $E_2^c$  were given in \eqref{structure.constants}.
Finally, the Hamiltonian of the model takes the simplest form in the basis of the gauge fields \cite{Balog:1993es,Sfetsos:2013wia,Bowcock:1988xr} 
\be
\label{Hamiltonian}
\begin{split}
&{\cal H}=\frac{k}{2\pi}\left(\tilde h_{AB} A_+^AA_+^B+h_{AB} A_-^AA_-^B\right)\,,\\
&\tilde h_{AB}=\Om_{AB}-(\l\Om\l^T)_{AB}\,,\quad h_{AB}=\Om_{AB}-(\Om\l^T\Om^{-1}\l\Om)_{AB}\,.
\end{split}
\ee
\paragraph{Isotropic case:}
For the deformation matrix $\l_A{}^B=\l\d_A{}^B$, the flat Lax connection was found in \eqref{ooskqsps}.  
The analysis of the Maillet brackets of the spatial component follows precisely that in \cite{Itsios:2014vfa} for the isotropic deformation 
of a general semi-simple group with $\Omega_{AB}=\delta_{AB}$. Skipping the details the end result is
\be
\begin{split}
&r_{+zw}=-\frac{2e^2(1+x+z^2(1-x))zw}{(z-w)(1-z^2)}\,\Om^{AB} T_A\otimes T_B\,,\quad r_{-zw}+r_{+wz}=0\,,\\
&x=\frac{1+\l^2}{2\l}\,,\quad e=\frac{2\l}{\sqrt{k(1-\l)(1+\l)^3}}\,,
\end{split}
\ee
satisfying the two consistency equations
\be
\label{cdkjssl}
r_{+zw}+r_{-wz}=0\,,\quad
r_{+z_2z_3}r_{+z_1z_2}=r_{+z_1z_3}r_{-z_1z_2}+r_{+z_2z_3}r_{+z_1z_3}\,.
\ee
The results are precisely those in \cite{Itsios:2014vfa}, and apply to a general semi-simple group or non-semisimple group, equipped with an $\Omega_{AB}$ which obeys \eqref{Omega.properties}.
\paragraph{Non-invertible $\l$-matrix:}For the deformation matrix $\l_A{}^B=\text{diag}(\l,-\l,0,0)$ the Lax connection was found in~\eqref{LAX.singular}. Then we may compute the spatial component ${\cal L}_1={\cal L}_1^AT_A$ expressed in terms of the currents ${\cal S}^A_\pm$ \eqref{djskwjdk}
\be
\begin{split}
&{\cal L}^1_1=-\frac{\sqrt{2\l}}{1-\l^2}\left((\nicefrac{\l}{z}-z){\cal S}_+^1-(\l z-\nicefrac1z){\cal S}_-^1\right)\,,\\
&{\cal L}^2_1=-\frac{\sqrt{2\l}}{1-\l^2}\left((\nicefrac{\l}{z}-z){\cal S}_+^2+(\l z-\nicefrac1z){\cal S}_-^2\right)\,,\\
&{\cal L}^3_1=-\frac{\l}{1-\l^2}\left((\l-z^2){\cal S}_+^3+(\l-\nicefrac{1}{z^2}){\cal S}_-^3\right)\,,\quad
{\cal L}^4_1={\cal S}_+^4+{\cal S}_-^4\,.
\end{split}
\ee
Employing the latter, \eqref{djskwjdk0} and \eqref{structure.constants}, we can compute the Poisson brackets of ${\cal L}_1$ which take the Maillet form \eqref{PB.spatial}, where the non-vanishing components $r_{+zw}^{AB}$ read
\be
\label{rpm.non.invertible}
\begin{split}
&r_{+zw}^{11}=r_{+zw}^{22}=\frac{4w(z^2-\l)(1-\l z^2)}{k(1-\l^2)z(z^2-w^2)}\,,\quad
r_{+zw}^{34}=\frac{2(z^2+w^2)(z^2-\l)(1-\l z^2)}{k(1-\l^2)z^2(z^2-w^2)}\,,\\
&r_{+zw}^{43}=r_{+zw}^{34}-\frac{2\l(1-z^4)}{kz^2(1-\l^2)}\, .
\end{split}
\ee
As a consistency check the $r_{-zw}^{AB}$ components satisfy the consistency properties
\be
r^{AB}_{+zw}+r^{BA}_{-wz}=0\,,\quad
r_{+z_1z_3}^{DC}r_{-z_1z_2}^{EB}\,f_{DE}{}^A+r_{+z_2z_3}^{DC}r_{+z_1z_2}^{AE}f_{DE}{}^B+r_{+z_2z_3}^{BD}r_{+z_1z_3}^{AE}f_{DE}{}^C=0
\ee
and for an isotropic deformation $r_{\pm zw}^{AB}=r_{\pm zw}\,\Om^{AB}$ these are in agreement with \eqref{cdkjssl}.
It would be worth investigating the existence of a description in terms of a twist function~\cite{Vicedo:2010qd}.

\subsection{Symmetry algebras}
\label{Sec:Symmetry}

\paragraph{Isotropic case:} Specializing to the $E_2^c$ case \eqref{Ec2.commutators} and for a deformation matrix $\l_A{}^B=\l\d_A{}^B$, we express the current algebra \eqref{djskwjdk0} in terms of the gauge fields $A^A_\pm$ using \eqref{djskwjdk}
\be
\label{Bowcock1}
{\cal S}^A_\pm=A^A_\pm-\l A^A_\mp\,,
\ee
yielding the non-vanishing Poisson brackets
\be
\label{skdmdkq}
\begin{split}
&\{A_\pm^A,A_\pm^B\}_\text{P.B.}=\frac{2f^{AB}{}_C}{k(1-\l)(1+\l)^2}\left((1+\l+\l^2)A_\pm^C-\l A_\mp^C\right)\d_{12}
\pm\frac2k\frac{\Om^{AB}}{1-\l^2}\d'_{12}\,,\\
&\{A_+^A,A_-^B\}_\text{P.B.}=\frac{2\l f^{AB}{}_C}{k(1-\l)(1+\l)^2}(A_+^C+A_-^C)\d_{12}\,,
\end{split}
\ee
where the $f^{AB}{}_C$'s of $E_2^c$ can be found in \eqref{structure.constants}. Two comments are in order concerning the above algebra. Firstly, as in the line element \eqref{gpp.su21} we can absorb the $b$ appearing in the element $\Om^{44}=-b$ \eqref{Omega.Inv}, by shifting $A^4_\pm$ as follows
\be
\label{shift.absorb}
A^4_\pm\to A^4_\pm+\frac{b}{2}A_\pm^3\,.
\ee
This is consistent with the fact that $b$ may and was also absorbed in the corresponding backgrounds as well.
Secondly, the dependence on $k$ is fictitious and it can be easily absorbed through  
\be
\label{k.dependence}
A^1_\pm\to\frac{A^1_\pm}{\sqrt{k}}\,,\quad A^2_\pm\to\frac{A^2_\pm}{\sqrt{k}}\,,\quad A^3_\pm\to A^3_\pm\,,\quad A^4_\pm\to \frac{A^4_\pm}{k}\,,
\ee
hence without loss of generality we take $k$ equal to one. Finally, we note that this cannot be done in the $SU(2)$ case where we consider in \eqref{skdmdkq} $\Om_{AB}=\d_{AB}$, $f^{AB}{}_C=\sqrt{2}\varepsilon_{ABC}$ and $T_A=\nicefrac{\s_A}{\sqrt{2}}$, with $A=1,2,3$.

\paragraph{Non-invertible $\l$-matrix:} Focusing to the deformation matrix $\l_A{}^B=\text{diag}(\l,-\l,0,0)$ for the $E_2^c$ case \eqref{Ec2.commutators}
we express the current algebra \eqref{djskwjdk0} in terms of the gauge fields $A^A_\pm$, as these are given in terms of the ${\cal S}_\pm$'s 
in \eqref{djskwjdk}
\be
\label{Bowcock2}
{\cal S}_\pm^1=A_\pm^1-\l A_\mp^1\,,\quad {\cal S}_\pm^2=A_\pm^2+\l A_\mp^2\,,\quad {\cal S}^3_\pm= A^3_\pm\,,\quad {\cal S}^4_\pm= A^4_\pm\,.
\ee
Employing the above and \eqref{structure.constants} yields the non-vanishing components
\be
\label{nsdsskls}
\begin{split}
&\{A^1_\pm,A_\pm^1\}_\text{P.B.}=\pm\frac2k\frac{\d'_{12}}{1-\l^2}\,,\quad \{A^2_\pm,A_\pm^2\}_\text{P.B.}=\pm\frac2k\frac{\d'_{12}}{1-\l^2}\,,\\ 
&\{A^3_\pm,A^4_\pm\}_\text{P.B.}=\pm\frac{2}{k}\d'_{12}\,,\quad
\{A^4_\pm,A^4_\pm\}_\text{P.B.}=\mp\frac{2b}{k}\d'_{12}\,,\\
&\{A_\pm^1,A_\pm^2\}_\text{P.B.}=\frac2k\frac{1}{(1-\l^2)^2}(A_\pm^3-\l^2A_\mp^3)\d_{12}\,,\\
&\{A_\pm^1,A_\pm^4\}_\text{P.B.}=-\frac2k\frac{1}{1-\l^2}(A_\pm^2+\l A_\mp^2)\d_{12}\,,\\
&\{A_\pm^2,A_\pm^4\}_\text{P.B.}=\frac2k\frac{1}{1-\l^2}(A_\pm^1-\l A_\mp^1)\d_{12}\,,\\
&\{A_+^1,A_-^2\}_\text{P.B.}=-\frac2k\frac{\l}{(1-\l^2)^2}(A_+^3-A_-^3)\d_{12}\,,\\
&\{A_+^1,A_-^4\}_\text{P.B.}=-\frac2k\frac{\l}{1-\l^2}(A_-^2+\l A_+^2)\d_{12}\,,\\
&\{A_+^2,A_-^4\}_\text{P.B.}=-\frac2k\frac{\l}{1-\l^2}(A_-^1-\l A_+^1)\d_{12}\,.
\end{split}
\ee
The parameters $b$ and $k$ appearing above can be set to zero and to one respectively, using again \eqref{shift.absorb} and \eqref{k.dependence}. 

\no
Interestingly, the algebra \eqref{nsdsskls} takes the form of \eqref{skdmdkq}. The change of basis can be easily found since they both result from \eqref{djskwjdk0}, \eqref{djskwjdk} and upon transforming \eqref{Bowcock2} to \eqref{Bowcock1} we find that
\be
\label{transform.As}
A_\pm^1\to A_\pm^1\,,\quad A^2_\pm\to\g(A^2_\pm-\b A^2_\mp)\,,\quad A^3_\pm\to A^3_\pm-\l A^3_\mp\,,\quad A^4_\pm\to A^4_\pm-\l A^4_\mp\ ,
\ee
where
\be
\b=\frac{2\l}{1+\l^2}\,,\quad \g=\frac{1}{\sqrt{1-\b^2}}\,.
\ee
The transformation  \eqref{transform.As} is a map from the singular to the isotropic deformation which preserves the Poisson brackets. However, it is not a canonical one, as the corresponding Hamiltonian densities \eqn{Hamiltonian} are not equal under \eqref{transform.As}.\footnote{In addition, we have checked that these are not equal up to total derivative term (with respect to $\s$). To check this we formed the difference $\d {\cal H}={\cal H}_\text{singular}-{\cal H}_\text{isotropic}$, and we used \eqref{transform.As} to express $\d {\cal H}$ in terms of the gauge fields of the isotropic deformation. Then, we inserted their explicit expressions using Eqs. \eqref{LR.E2C} and \eqref{D.E2C}. Finally, with the aid of footnote~\ref{worldsheet.der} in the resulting expression for $\d {\cal H}$,  we found terms which cannot be written as total derivative form with respect to $\s$, i.e., 
$$
\d {\cal H}=\frac{2\l(1+\l)}{(1-\l)^2}\del_\tau u\del_\tau v+\cdots
$$}

 \section{The contracted algebra} 
\label{Section.algebra}

In this Section we derive the symmetry algebras \eqref{skdmdkq} and \eqref{nsdsskls} of the isotropic and the non-invertible $\l$-deformed $E_2^c$ respectively, as a Saletan type of contraction~\cite{Saletan:1961} and as a Penrose limit of the symmetry algebra of the isotropic $\l$-deformed $SU(2)$~\cite{Balog:1993es,Sfetsos:2013wia,Georgiou:2016iom,Georgiou:2016zyo} times a timelike boson.

\subsection{The Saletan-type algebraic contraction}

Our starting point will be the symmetry algebra of the isotropic $\l$-deformed model~\eqref{lambda.action.singular} for the holomorphic and anti-holomorphic currents both at level $k$~\cite{Balog:1993es,Sfetsos:2013wia,Georgiou:2016iom}. In the early works this algebra was found by restricting its form with various consistency conditions as well as parity considerations. In contradistinction and more generally, we may consider the isotropic $\l$-deformed model at unequal levels for the holomorphic and anti-holomorphic currents at levels $k_L$ and $k_R$, respectively~\cite{Georgiou:2016zyo}. In this work the symmetry algebra at leading order in the $\nicefrac{1}{\sqrt{k_{L,R}}}$ expansion and exact in the isotropic deformation parameter $\lambda$ was computed from first principles from the corresponding two- and three-points functions of the current operators. Specifically,
\be
\label{OPEcurrents}
\begin{split}
&J^AJ^B=\frac{\Omega^{AB}}{x_{12}^{2}}+a_L\frac{if^{AB}{}_C\,J^C}{x_{12}}
+b_L\frac{if^{AB}{}_C\,\bar J^C \bar x_{12}}{x^2_{12}}+\dots\,,\\
&\bar J^A\bar J^A=\frac{\Omega^{AB}}{\bar x_{12}^{2}}+a_R\frac{if^{AB}{}_C\,\bar J^C}{\bar x_{12}}
+b_R\frac{if^{AB}{}_C\, J^C x_{12}}{\bar x^2_{12}}+\dots\,,\\
&J^A\bar J^B=b_R\frac{if^{AB}{}_C\,\bar J^C}{x_{12}}
+b_L\frac{if^{AB}{}_C\, J^C}{\bar x_{12}}+\dots\,,
\\
&
J_0 J_0 = -{1\ov x_{12}^2}\ ,\quad \bar J_0 \bar J_0 = -{1\ov \bar x_{12}^2}\, , 
\end{split}
\ee
where in the left-hand side the first operator is at $(x_1,\bar x_1)$ and the second at $(x_2,\bar x_2)$, whereas in the right-hand side the operator is always at $(x_2,\bar x_2)$. In addition, the last line represents the timelike boson and the various constants are given by
\be
\label{ablr}
\begin{split}
&a_L=\frac{1-\l_0\l^3}{\sqrt{k_L(1-\lambda^2)^3}}\ ,\qquad b_L={\l (\l_0^{-1}-\l)\ov \sqrt{k_R (1-\l^2)^3}}\ ,
\\
&a_R=\frac{1-\l_0^{-1}\l^3} {\sqrt{k_R(1-\lambda^2)^3}}\ , \qquad b_R={\l (\l_0-\l)\ov \sqrt{k_L (1-\l^2)^3}}\  ,
\\
&
\l_0=\sqrt{k_L\ov k_R}\ ,\qquad k=\sqrt{k_Lk_R}\ .
\end{split}
\ee
The above algebra is invariant under the parity-like transformation $J_a\leftrightarrow \bar J_a$, $x_i \leftrightarrow \bar x_i$ and $(a_L,b_L)\leftrightarrow (a_R,b_R)$ where the latter is implied by $k_L\leftrightarrow k_R$. Additionally, it is invariant under the non-perturbative symmetry   $k_L \to -k_R$ , $k_R\to -k_L$  and $\l\to  \nicefrac{1}{\l}$. We emphasize that the symmetry algebra of the action \eqn{lambda.action.singular} is the one given above in \eqn{OPEcurrents}, but with $k_L= k_R=k$.

\no
Specializing to the $SU(2)$ case, we have in our normalizations that $f^{AB}{}_C= \sqrt{2}\, \e_{ABC}\,,\,\, \Omega^{AB}=\delta_{AB}$ with $A=1,2,3$. In addition, we perform the basis change
\be
\label{kfjn}
\begin{split}
& J= \sqrt{k\ov 2} (J_3+J_0)\ ,\qquad F= \sqrt{1\ov 2 k} (J_3-J_0)\  ,
\\
&
J_3 =\sqrt{1\ov 2 k}\, J +\sqrt{k\ov 2}\, F \ ,\qquad  J_0 = \sqrt{1\ov 2 k}\, J -\sqrt{k\ov 2}\, F \ .
\end{split}
\ee
The above is inspired by the corresponding Penrose limit as we will see in the next subsection.
Then, in the limit $k_{L,R}\to \infty$ with $\l_0$ finite, we have the non-vanishing OPE's with 
\be
\label{contalg2}
\begin{split}
& JF= {1\ov x_{12}^2}\ , \\
&J J_i = i \a_L \e_{ij} {J_j\ov x_{12}}  + i \b_L \e_{ij} {\bar J_j \bar x_{12}\ov x_{12}^2}\,,\\
&J _iJ_j = {\d_{ij}\ov x_{12}^2} + i \a_L \e_{ij}  {F\ov x_{12}}  + i \b_L \e_{ij} {\bar F \bar x_{12}\ov x_{12}^2}\ ,
\\
&J \bar J_i = i \b_R \e_{ij} {\bar J_j\ov x_{12}}  + i \b_L \e_{ij} { J_j \ov \bar x_{12}} \ ,\\
& J_i \bar J_j =  i \b_R \e_{ij} {\bar F\ov x_{12}}  + i  \b_L \e_{ij} {F \ov \bar  x_{12}} \ ,
\end{split}
\ee
where $i,j=1,2$ and 
\be
\label{ablrg}
\begin{split}
&\a_L=\frac{\l_0^{-\nicefrac12}(1-\l_0\l^3)}{(1-\lambda^2)^{\nicefrac32}}\ ,\qquad \b_L={\l_0^{\nicefrac12}\l (\l_0^{-1}-\l)\ov (1-\l^2)^{\nicefrac32}}\ ,
\\
&\a_R=\frac{\l_0^{\nicefrac12}(1-\l_0^{-1}\l^3)} {(1-\lambda)^{\nicefrac32}}\ , \qquad \b_R={\l_0^{-\nicefrac12} \l (\l_0-\l)\ov (1-\l^2)^{\nicefrac32}}\ ,
\end{split}
\ee
with the rest of the OPE's obtained using parity.
For the case $\l_0=1$ we have that 
\be
\label{ablr}
\a_L= \a_R=\frac{1+\l +\l^2}{\sqrt{(1-\lambda)(1+\l)^3}}\ ,\qquad 
\b_L=\b_R=\frac{\l}{\sqrt{(1-\lambda)(1+\l)^3}}\, .
\ee

\no
The following question arises. In \eqn{kfjn} we have combined the two operators $J_3$ and $\bar J_3$ which only at the CFT point, that is at $\l=0$, 
 have dimension one. For $\l\neq 0$ they acquire anomalous dimensions given by 
\be
\g_L= {4 \ov k_R} {\l^2 (\l-\l_0^{-1})^2\ov (1-\l^2)^3}\ ,\qq \g_R= {4 \ov k_L} {\l^2 (\l-\l_0)^2\ov (1-\l^2)^3}\,.
\ee
Their two-point functions  up to ${\cal O}(\nicefrac1k)$ are given by
\be
\label{jj33}
J_3 J_3 = {1\ov x_{12}^2}\Big(1 + \g_L \ln {\varepsilon^2\ov |x_{12}|^2}\Big)\,,\quad J_3\bar J_3 = - {\sqrt{\g_L \g_R} \ov |x_{12}|^2}\,,
\ee
where $\varepsilon$ is a short distance regulator. The first of the OPE's above was computed in 
 \cite{Georgiou:2016zyo}, whereas the second OPE one can be found computed the lines of~\cite{Georgiou:2016iom}.

\no
Then, using \eqn{kfjn} we have that
\be
\label{sksfkjslq}
J J = {\tilde \g_L\ov x_{12}^2}  \ln {\varepsilon^2\ov |x_{12} |^2}\,,\quad 
\bar J \bar J = {\tilde \g_R\ov x_{12}^2}  \ln {\varepsilon^2\ov |x_{12} |^2}\,,\quad  J\bar J = - {\d \ov |x_{12}|^2}\  .
\ee
Note that, even though is a bit unconventional, we retain in the OPE's the cutoff $\e$. The various parameters are given by
\be 
\begin{split}
&  \tilde \g_L = 2\l_0  {\l^2 (\l-\l_0^{-1})^2\ov (1-\l^2)^3} \ ,\qq \tilde \g_R = 2\l_0^{-1}  {\l^2 (\l-\l_0)^2\ov (1-\l^2)^3} \ ,
 \\
 & \d =  \sqrt{\tilde \g_L \tilde \g_R} = 2   {\l^2 (\l-\l_0) (\l-\l_0^{-1})\ov (1-\l^2)^3} \ .
 \end{split}
\ee
For $\l=0$ we have $\b_{L,R}=0$ and $\a_{L,R}=1$ so that the algebra becomes 
 \be
\begin{split}
& JF= {1\ov x_{12}^2}\ , \qq
J J_i = i  \e_{ij} {J_j\ov x_{12}}  \ ,\qq
J _iJ_j = {\d_{ij}\ov x_{12}^2} + i \e_{ij}  {F\ov x_{12}}  \ ,
\end{split}
\ee
plus an identical copy for the anti-holomorphic part, which is the OPE for the $E_2^c$ generators. Note that, compared to the generators of Section~\ref{Sec2}  the $J_i$'s  correspond to the $i P_i$'s and $(J,F)$ to $(i J,i F)$. 

\no
Finally, let us make the connection of the OPE in \eqref{contalg2} and the current algebra of the isotropic $\l$-deformed $E_2^c$ model in \eqref{skdmdkq} for $\l_0=1$. It can be shown that \eqref{skdmdkq} can be derived from \eqref{contalg2}, using the point splitting procedure prescribed in Section 6 of~\cite{Georgiou:2016iom}, upon taking the classical limit $[\cdot,\cdot]\to -i\pi\{\cdot,\cdot\}_\text{P.B.}$ and also appropriately rescaling the currents
\be
J^A\to-\frac{iA_+^A}{\sqrt{1-\l^2}}\ ,\qquad \bar J^A\to-\frac{iA_-^A}{\sqrt{1-\l^2}}\ .
\ee
In addition, applying  above procedure in \eqref{sksfkjslq} yields
\be
\{J(\s_1),J(\s_2)\}_\text{P.B.}=2\tilde\g_L\d'(\s_{12})\ln\frac{\varepsilon^2}{\s_{12}^2}\,,\quad
\{J(\s_1),\bar J(\s_2)\}_\text{P.B.}=0\,,
\ee
It can be easily see that the Jacobi identity for the Poisson bracket is not spoiled by these additional brackets.

\no
We may argue that the OPE algebra \eqref{contalg2} enhanced with \eqref{sksfkjslq} is in fact exact. Recall that, the dots in the OPE's  \eqref{OPEcurrents} represent 
terms of ${\cal O}(\nicefrac{1}{k^{\nicefrac{3}{2}}})$. Since the basis change \eqref{kfjn} has terms of relative order $\nicefrac{1}{k}$, the terms of ${\cal O}(\nicefrac{1}{k^{\nicefrac{3}{2}}})$ in \eqref{OPEcurrents} are completely washed out in the $k\to \infty$ limit.

\subsection{Contraction induced by the Penrose limit}

The Penrose type of limit \eqn{pensorel} should also induce a Saletan-type of contraction in the WZW currents and their $\l$-deformed versions~\cite{Balog:1993es,Sfetsos:2013wia}. To see that we first parametrize the $SU(2)$ group element as in the $\l$-deformed model $SU(2)$ \eqref{g.su2}, \eqref{B.su2}~\cite{Sfetsos:2013wia}
\be
g=e^{i \a_a \s_a}=
\begin{pmatrix}
 \cos\a + i \sin\a \cos\b & \sin\a \sin\b\, e^{-i\g} \\
-\sin\a \sin\b\, e^{i\g} & \cos \a  - i \sin\a \cos\b 
\end{pmatrix}\ ,
\label{su2g2}
\ee
where 
\be
\a_1 = -\a \sin\b \sin\g \ ,\qq  \a_2 = \a \sin\b \cos\g\ ,\qq   \a_3 = \a \cos\b\,.
\ee
Then, the right-invariant Maurer--Cartan one-forms $R=R^AT_A$ for  $T_A=\nicefrac{1}{\sqrt{2}}\,\s_A$, $\Om_{AB}=\d_{AB}$ with $A=1,2,3$ are found through \eqref{RL.MC.forms} 
\be
\label{Rforms}
\begin{split}
R_\pm&=-\,e^{i\g}\left(\mp i\sin\b\, d\a-\sin\a(\pm i\cos\a\cos\b+\sin\a)d\b\right.\\
&\left.+\sin\a\sin\b(\cos\a\mp i\sin\a\cos\b)d\g\right)\,,\\
R_3&=-\sqrt{2}\left(-\cos\b\,d\a+ \sin \a \cos \a\sin\b \, d\b + \sin^2 \a \sin^2 \b \, d\g\right)\,,
\end{split}
\ee
where we denote $R_\pm = \nicefrac{1}{\sqrt{2}}(R_1\pm i R_2)$. 
In addition, the left-invariant Maurer--Cartan one-forms $L=L^AT_A$ can be found from \eqref{RL.MC.forms} 
\be
\label{Lforms}
\begin{split}
&L_\pm = \, e^{\pm i \g}\Big(\pm i \sin\b\ d\a + \sin\a(\pm i \cos \a \cos\b-\sin\a)\ d\b
\\
& \qq -\sin\a \sin\b (\cos\a \pm i\sin\a \cos \b)\ d\g \Big)\ ,
\\
&L_3  =  \sqrt{2  } \left(\cos\b \ d\a - \sin \a \cos \a\sin\b \, d\b + \sin^2 \a \sin^2 \b \, d\g\right)\ ,
\end{split}
\ee
where we denote $L_\pm = \nicefrac{1}{\sqrt{2}} (L_1\pm i L_2)$. In addition, the adjoint action \eqref{RL.MC.forms} is given by
\be
\begin{split}
&D_{11}=\cos^2\a-\sin^2\a(\cos^2\b+\sin^2\b\cos2\g)\,,\\ 
&D_{12}=\sin2\a\cos\b-\sin^2\a\sin^2\b\sin2\g\,,\\
&D_{13}=-\sin2\a\sin\b\cos\g-\sin^2\a\sin2\b\sin\g\,,\\
&D_{21}=-\sin2\a\cos\b-\sin^2\a\sin^2\b\sin2\g\,,\\
&D_{22}=\cos^2\a-\sin^2\a(\cos^2\b-\sin^2\b\cos2\g)\,,\\
&D_{23}=\sin^2\a\sin2\b\cos\g-\sin2\a\sin\b\sin\g\,,\\
&D_{31}=\sin2\a\sin\b\cos\g-\sin^2\a\sin2\b\sin\g\,,\\
&D_{32}=\sin^2\a\sin2\b\cos\g+\sin2\a\sin\b\sin\g\,,\\
&D_{33}=\cos^2\a+\sin^2\a\cos2\b
\end{split}
\ee
and it satisfies the identities
\be
R_A=D_{AB}L_B\,,\quad D_{AC}D_{BC}=\d_{AB}\,.
\ee
Let us now consider the $\l$-deformed left- and right-invariant $SU(2)$ Maurer--Cartan currents which are given in terms of~\cite{Balog:1993es,Sfetsos:2013wia,Hollowood:2014rla}
\be\
\label{sknckmslq}
{\cal J}=i \sqrt{k}(\mathbb{I}-\l D)^{-1} \del_+g g^{-1}\,,\quad  \bar{\cal J}=-i \sqrt{k}(\mathbb{I}-\l D^T)^{-1} g^{-1}\del_-g\,,
\ee
and also introduce the currents for the timelike direction $t$
\be
\label{sknckmslq1}
{\cal J}_0=-\frac{\sqrt{2k}}{1+\l}\,\del_+t\,,\quad \bar {\cal J}_0=-\frac{\sqrt{2k}}{1+\l}\,\del_-t\,.
\ee
Upon applying the pp-wave limit of \eqref{pensorel} in \eqref{sknckmslq} and \eqref{sknckmslq1}, we obtain the  dressed currents of the non-invertible $\l$-deformed $E_2^c$
\be
\begin{split}
J_\pm&=\frac{1}{\sqrt{2}}({\cal J}_1\pm i{\cal J}_2)\\
&=\frac{1}{\sqrt{2}(1-\l^2)^{\nicefrac32}}\text{e}^{\pm\frac{x^+}{2}\sqrt{\frac{1+\l}{1-\l}}}
\left(\mp i((1+\l)^2\del_+x_1\mp i(1-\l^2)\del_+x_2)\right.\\
&\left.-\frac{1}{2}\sqrt{\frac{1+\l}{1-\l}}((1-\l)^2x_1\mp i(1-\l^2)x_2)\del_+x^+\right),\\
J&=\sqrt{\frac{k}{2}}\left({\cal J}_3+ {\cal J}_0\right)\\
&=\frac{1}{\sqrt{1-\l^2}}\left(\del_+x^--\frac14\left(\frac{1-\l}{1+\l}x_1^2+\frac{1+\l}{1-\l}x_2^2\right)\del_+x^+\right)\\
&-\frac{1}{2(1-\l)}\left(\frac{1-\l}{1+\l}x_1\del_+x_2-\frac{1+\l}{1-\l}x_2\del_+x_1\right),\\
F&=\sqrt{\frac{1}{2k}}\left({\cal J}_3-{\cal J}_0\right)=\frac{\del_+x^+}{\sqrt{1-\l^2}}
\end{split}
\ee
and
\be
\begin{split}
\bar J_\pm&=\frac{1}{\sqrt{2}}(\bar{\cal J}_1\pm i\bar{\cal J}_2)\\
&=\frac{1}{\sqrt{2}(1-\l^2)^{\nicefrac32}}\text{e}^{\pm\frac{x^+}{2}\sqrt{\frac{1+\l}{1-\l}}}
\left(\pm i((1+\l)^2\del_-x_1\pm i(1-\l^2)\del_-x_2)\right.\\
&\left.+\frac{1}{2}\sqrt{\frac{1+\l}{1-\l}}((1-\l)^2x_1\pm i(1-\l^2)x_2)\del_-x^+\right),\\
\bar J&=\sqrt{\frac{k}{2}}\left(\bar{\cal J}_3+\bar {\cal J}_0\right)\\
&=\frac{1}{\sqrt{1-\l^2}}\left(\del_-x^--\frac14\left(\frac{1-\l}{1+\l}x_1^2+\frac{1+\l}{1-\l}x_2^2\right)\del_-x^+\right)\\
&+\frac{1}{2(1-\l)}\left(\frac{1-\l}{1+\l}x_1\del_-x_2-\frac{1+\l}{1-\l}x_2\del_-x_1\right),\\
\bar F&=\sqrt{\frac{1}{2k}}\left(\bar{\cal J}_3-\bar {\cal J}_0\right)=\frac{\del_-x^+}{\sqrt{1-\l^2}}\, .
\end{split}
\ee
Hence, we have explicitly demonstrated that the Penrose limit  on the gravitational background is in one to one correspondence with the Saletan-type contraction of the corresponding symmetry algebra.

\section{Conclusion and Outlook}
\label{Sec5}

In this work we studied $\s$-models obtained from $\l$-deformations of WZW models on the non-semi-simple group $E_2^c$, i.e. the central extension of the Euclidean group $E_2$. In particular, for an isotropic and a non-invertible deformation matrix $\l_A{}^B$, the target spacetimes assume the form of a plane wave metric and have a non-trivial  antisymmetric tensor. In addition, these models may be obtained as appropriate Penrose limits of the $\l$-deformed $SU(2)$ background times a timelike scalar.

\no
Furthermore, these two deformations turn out to integrable in the Hamiltonian sense as their equations of motion take the form of a flat Lax connection and the corresponding conserved charges are independent. We have also studied the corresponding symmetry algebras which turn out to be the same but the Hamiltonians differ. Thus, they are not related via a canonical transformation.

\no
Finally, we worked out the contracted algebras using two distinct approaches. Firstly, we considered a Saletan-type of contraction on the algebra of the 
isotropic $\l$-deformed $SU(2)$ times a timelike scalar. Secondly, we applied a Penrose limit on the $\l$-deformed currents which obeys the algebra of the aforementioned model. In fact the Penrose limit induces the deformation of the symmetry algebra.

\no
An extension of the present work is to consider the (bi-)Yang--Baxter integrable deformations~\cite{Klimcik:2002zj,Klimcik:2008eq,Klimcik:2014} of the PCM for the non-semi-simple group $E_2^c$. The starting point for this analysis would be to find solutions of the modified classical Yang--Baxter equation for the corresponding matrix ${\cal R}^A{}_B$ which is antisymmetric, when both indices are lowered or raised using \eqref{Omega.Inv}. The corresponding model is expected to be related to the isotropic $\l$-matrix on $E_2^c$, studied in Section~\ref{isotropic.section}, up to an analytic continuation and a Poisson--Lie T-duality~\cite{Vicedo:2015pna,Hoare:2015gda,Sfetsos:2015nya,Klimcik:2015gba,Klimcik:2016rov}.

\subsection*{Acknowledgements}

This research work was supported by the Hellenic Foundation for Research and Innovation (H.F.R.I.) under the ``First Call for H.F.R.I. Research Projects to support Faculty members and Researchers and the procurement of high-cost research equipment grant'' (MIS 1857, Project Number: 16519).

 \appendix

 \section{Plane wave limits}\label{AppA}
 
 In this Appendix we will consider various Penrose type of limits in the target spacetime of integrable $\s$-models.
 
 \subsection{The plane wave of the $\l$-deformed models}

The examples involving $\l$-deformed models are the most interesting ones since the deformation parameters are still present non-trivially after the Penrose limit is taken.

\subsubsection{The $\l$-deformed version of the bi--Yang--Baxter model}
\label{Sec:lambdaYB}

In~\cite{Sfetsos:2015nya} and specializing to the $SU(2)$ case a two-parameter $\l$-deformation was constructed as the Poisson--Lie T-dual,
accompanied by complexification, of the bi-Yang--Baxter model~\cite{Klimcik:2008eq,Klimcik:2014}. Specifically, the line element and the field strength of the antisymmetric tensor were given in Eq.(6.1) of~\cite{Sfetsos:2015nya}. In the normalization of~\eqref{action} they read
\be
\label{lYBg}
\begin{split}
ds^2 = &  2 k\left\{ \frac{1+\l}{1-\l} \left(1 + \zeta^2 \frac{(1+\l)^2 }{\D(\a,\b)}  \sin^2\a \sin^2\b \right) d\a^2\right.
\\
&\left. + \frac{1-\l^2}{\D(\a,\b)} \sin^2\a  \Big(d\beta^2+\sin^2\beta\, d\gamma^2\Big)
+ 2\zeta  \frac{(1+\l)^2 }{\D(\a,\b)}  \sin^2\a \sin \b\, d\a\,d\b  \right\}
  \end{split}
  \ee
   and 
   \be
   \label{lYBB}
   \begin{split}
H  =  &  - {2 k \ov \D^2(\a,\b)}  \bigg\{  \Big(4\l - \zeta^2(1+\lambda)^2\Big)\Delta(\a,\b)
\\
&  \ +2  (1-\l^2)^2 + 2 \zeta^2 (1+\l)^4 \sin^2\a \sin^2\b   \bigg\}   \,
  \sin^2\alpha\,\sin\beta\,  d\alpha \wedge \,d \beta \wedge d\g\,, 
  \end{split}
  \ee
  where
   \be
   \Delta(\a,\b)  =  1+ \l^2 - 2\l \cos2\a + \zeta (\l^2-1)  \sin 2\a \cos\beta + \zeta^2 (1+\l)^2 \sin^2\a \cos^2\b\, .       
\end{equation}
For $\zeta=0$, the latter model simplifies to the $\l$-deformed $SU(2)$ \eqref{g.su2}, \eqref{B.su2}.

\no
We perform the limit \eqn{pensorel} since \eqn{geo} is still a geodesic (we also let in the Cartesian coordinates $x_2\to x_2 - \zeta x_1)$.  We find a 
non-conformally flat (non-vanishing Weyl tensor of the Petrov type N) plane wave background for non-vanishing $\zeta$ or $\l$ with line element
 \be
\label{ppwaltez}
 ds^{2}  = 2 d  x^+ d x^- + dx_1^2 +  dx_2^2 + F_{ij}\,x_i x_j   (dx^+)^2 \ ,
\ee
where $F_{ij}$ is a negative definite symmetric matrix with elements
\be
\begin{split}
&  F_{11} = - {1\ov 4} \Big({1-\l\ov 1+\l}\Big)^2 - {\zeta^2 \ov 4}\, {3 \l^2 - 2 \l + 3\ov (1-\l)^2}  - {\zeta^4 \ov 4}\, \left({1+\l\ov 1-\l}\right)^2\ ,
\\
& F_{22} = - {1+\zeta^2\ov 4} \left({1+\l\ov 1-\l}\right)^2\ ,\quad
F_{12} = {\zeta \ov 4}\, { 2 (1+\l^2) + \zeta^2 (1+\l)^2\ov (1-\l)^2}\ .
\end{split}
\ee
In addition, the field strength of the antisymmetric tensor equals to
\be
H = -{2 (1+\l^2) + \zeta^2 (1+\l)^2\ov 2 (1-\l^2)} \,   dx^+\wedge dx_1 \wedge dx_2\,.
\ee
Note that these expressions for $\zeta=0$ truncate to the ones in \eqref{metric.singular} and \eqref{H.singular}.

\subsubsection{The $\l$-deformed $\nicefrac{SU(2)}{U(1)}$ }

We will consider the $\l$-model on symmetric coset spaces constructed in~\cite{Sfetsos:2013wia} and whose integrability was shown in~\cite{Hollowood:2014rla}. Specifically, we consider the $\nicefrac{SU(2)}{U(1)}$ case whose line element in the parameterization of Eq.(4.1) in~\cite{Itsios:2014lca} (in the normalization of \eqref{action}) reads
\be
ds^2=2k\left\{\frac{1-\l}{1+\l}(d\th^2+\tan^2\th\,d\phi^2)+\frac{4\l}{1-\l^2}\left(\cos\phi d\th-\tan\th\sin\phi\,d\phi\right)^2\right\}\,.
\ee
Adding to the above the term $-2k\,dt^2$ and considering the null geodesic Penrose limit
\be
\th=u\,,\quad t=\sqrt{\frac{1+\l}{1-\l}}u-\frac{1}{2k}\sqrt{\frac{1-\l}{1+\l}}v\,,\quad \phi=\frac{1}{\sqrt{2k}}\sqrt{\frac{1+\l}{1-\l}}\,x\,,\quad k\to \infty\,,
\ee
leads to the target spacetime 
\be
\label{baksfe}
ds^2=2dudv+\tan^2u dx^2-\frac{4\l }{(1-\l)^2}(x^2du+2x\tan u dx ) du\ .
\ee
 It can be easily checked that the above line element describes a conformally flat (vanishing Cotton tensor) plane wave, where $\xi=\del_v$ is a null and geodesic vector, and has vanishing scalar invariants. In addition, \eqn{baksfe} may be considered as  the deformation of the plane-wave solution found in Eq.(1.3) of \cite{Bakas:2002qh} via a contraction procedure involving the $\nicefrac{SU(2)}{U(1)}$ coset CFT parafermions and a time-like boson, leading to a logarithmic CFT. It would be worth exploring the nature of the aforementioned marginal deformation.\footnote{If we wish, we may preserve conformality by introducing the dilaton
\begin{equation*}
\Phi=-\ln\cos u+\frac{\l u^2}{(1-\l)^2}\,.
\end{equation*}} Finally, we can bring the above line element into the Brinkmann form using the following transformation
\be
x\to\frac{x}{\tan u}\,,\quad v\to v+\frac{x^2}{\sin2u}+\frac{2\l x^2}{(1-\l)^2\tan u}\,,
\ee
leading to 
\be
ds^2=2dudv+dx^2+F(u)x^2du^2\,,\quad F(u)=2\left(\frac{2\l}{(1-\l)^2}+\frac{1}{\cos^2u}\right)\,.
\ee

 \subsection{The plane wave of Yang--Baxter models}

In the examples we present below the Yang--Baxter models are involved. It turns out that the deformation parameters effectively drop out  after the Penrose limit is taken.
 
\subsubsection{The bi-Yang--Baxter model}
\label{Section.biYB}

Let us now consider the bi-Yang--Baxter model constructed in~\cite{Klimcik:2008eq} and its Lax connection was found in~\cite{Klimcik:2014}.

\no
The bi-Yang--Baxter $\sigma$-model in the $SU(2)$ case can be found in Eq.(6.1) of~\cite{Sfetsos:2015nya} and in the normalization of \eqref{action} the line element reads
\be
\label{etasu}
\begin{split}
ds^2 &=  \frac{2}{ T\Lambda} \Big(  d\theta^2  + d \psi^2+ d\phi^2+
2\cos\theta d\psi d\varphi\\ 
&\qq\qq\qq\qq  + \big((\zeta + \eta \cos\theta) d \psi +(\eta + \zeta \cos\theta)d\varphi\big)^2  \Big) \ ,
\\
\Lambda &=  1+\eta^2 +\zeta^2 + 2 \eta\,\zeta \cos\theta \ ,
\end{split}
\ee
whereas the antisymmetric tensor vanishes.
It is convenient to change variables as $\th =2 \om$, $\psi= \phi_1+\phi_2 $ and $\phi= \phi_1-\phi_2 $. 
Then
\be
\label{etasu2}
\begin{split}
ds^2 &=  \frac{8}{ T \Lambda} \Big(  d\om^2  + \cos^2\om\, d \phi_1^2+ \sin^2\om\, d \phi_2^2 \\ 
&\qq\qq\qq\qq + \big( (\zeta + \eta) \cos^2 \om\, d\phi_1  +
  (\zeta -\eta) \sin^2 \om \, d\phi_2\big)^2  \Big) \ ,
\\
\Lambda &=  1+ (\zeta+\eta) ^2\cos^2\om  +(\zeta-\eta)^2\sin^2\om\ .
\end{split}
\ee
Note that up to an analytic continuation and a Poisson--Lie T-duality~\cite{Vicedo:2015pna,Hoare:2015gda,Sfetsos:2015nya,Klimcik:2015gba,Klimcik:2016rov},
it is related to the bi-Yang--Baxter $\s$-model on $SU(2)$, constructed in the Section~\ref{Sec:lambdaYB}.

\no
Next we add the term $-\nicefrac8T\, dt^2$ into the line element \eqref{etasu2}. The null geodesic is  $t-\phi_1=0$, $\om=0$. Let ($k=\nicefrac4T$)
\be
\begin{split}
t= {1\ov 2} \Big(x^+ - {x^-\ov k}\Big)\ ,
\quad \phi_1 = {1\ov 2} \Big(x^+ + {x^-\ov k}\Big)\,,\quad
\om = \sqrt{1+(\zeta + \eta)^2}\, {\r\ov \sqrt{2k}}\ .
\end{split}
\ee
Then, in the limit $k\to \infty$ we obtain a conformally flat plane wave background with line element
\be
ds^2 = 2 dx^+ dx^- + d\r^2 + \r^2 d\phi_2^2  -\frac14\Big(1 + 2(\eta^2  +\zeta^2)\Big) \r^2\,  (dx^+)^2 -  (\eta^2-\zeta^2)\r^2 d\phi_2dx^+\ .
\ee
Letting $\phi_2\to \phi_2 +\frac12 (\eta^2-\zeta^2) x^+$ we find\footnote{This shift implies for the original angles 
\begin{equation*}
\psi = {1-\zeta^2+\eta^2\ov 2} x^+ + {x^-\ov 2k}+ \phi_2 \,, \quad \phi = {1+\zeta^2-\eta^2\ov 2} x^+ + {x^-\ov 2k}- \phi_2\,.
\end{equation*}
}
\be
ds^2 = 2 dx^+ dx^- + d\r^2 + \r^2 d\phi_2^2  -\frac14 \Big(1+ (\zeta + \eta)^2\Big) \Big(1+ (\zeta - \eta)^2\Big)\, \r^2\,  (dx^+)^2\,,
\ee
or expressed in Cartesian coordinates
\be
\label{biYB.PCM}
ds^2 = 2 dx^+ dx^- + dx_1^2 + dx_2^2   -{x_1^2+x_2^2\ov 4} \Big(1+ (\zeta + \eta)^2\Big) \Big(1+ (\zeta - \eta)^2\Big)\, \,  (dx^+)^2 \ .
\ee
Note that with an appropriate rescaling of the light-cone coordinates $x^\pm$ we can absorb the $\eta$ and $\zeta$ dependence
\be
\label{PCM}
ds^2 = 2 dx^+ dx^- + dx_1^2 + dx_2^2   -{x_1^2+x_2^2\ov 4} \,  (dx^+)^2\,,
\ee
corresponding to the plane wave limit of the (isotropic) PCM 
$
{\cal L}_\text{PCM}=\kappa^2 L^A_+L^A_-\,,
$ 
for the $SU(2)$ times a timelike coordinate.

\no
Alternatively, it can be directly obtained from the Lagrangian density of the isotropic PCM 
$
{\cal L}_\text{PCM}=\kappa^2\Om_{AB}L^A_+L^B_-\,,
$
on the group $E_2^c$. Employing \eqref{RL.MC.forms}, \eqref{Omega.Inv} and \eqref{LR.E2C}, as well as
the coordinate transformations (a variant of \eqref{atox0})
\be
a_1= x_2 + x_1 \cos u\, , \quad a_2= x_1 \sin u\, ,
\quad v\to v + \frac12x_1 x_2 \sin u\, ,
\ee
and \eqref{rotate12}, we also use
\be
\begin{split}
&x_1\to\frac{x_1}{\sqrt{2}\cos\frac{x^+}{2}}\,,\quad x_2\to\frac{x_2}{\sqrt{2}\sin\frac{x^+}{2}}\,,\quad u\to x^+\,,\\ 
&v\to x^--\frac{b}{2} x^+-\frac14\left(x_1^2\tan\frac{x^+}{2}-x_2^2\cot\frac{x^+}{2}\right)\,,
\end{split}
\ee
reaching the line element \eqref{PCM}, in the normalization of \eqref{action}.

\subsubsection{The $\eta$-deformed $\nicefrac{SU(2)}{U(1)}$}

We consider the integrable Yang--Baxter $\s$-model on symmetric coset spaces constructed in~\cite{Delduc:2013fga}. Specifically, we will consider the $\nicefrac{SU(2)}{U(1)}$ case, whose line element can be read from Eq.(4.2) of~\cite{Delduc:2013fga}  and in the normalization of \eqref{action} it reads\footnote{Upon applying stereographic coordinates  
$
z=\cot\frac{\theta}{2}e^{i\phi}\,,
$
with $T=\frac{1}{4\zeta}$.
}
\be
ds^2=T\, \frac{1+\eta^2}{1+\eta^2\cos^2\th}\, (d\th^2+\sin^2\th d\phi^2)\,.
\ee
Adding to the above line element the term $-T\,dt^2$ and considering the null geodesic Penrose limits for $T\to\infty$
\be
t=\sqrt{1+\eta^2}\,x^+-\frac{1}{\sqrt{1+\eta^2}}\frac{x^-}{T}\,,\quad \th=\frac{\pi}{2}-\frac{x}{\sqrt{T}}\,,\quad \phi=x^+\,,
\ee
or
\be
t=\sqrt{1+\eta^2}\,x^+-\frac{1}{2\sqrt{1+\eta^2}}\frac{x^-}{T}\,,\quad 
\th=\frac{\pi}{2}-\frac{x}{\sqrt{T}}\,,\quad
\phi= x^++\frac{1}{2(1+\eta^2)}\frac{x^-}{T}\,,
\ee
leads to a conformally flat plane (vanishing Cotton tensor) wave background with the same line element expressed in Brinkmann coordinates
\be
ds^2=2dx^+dx^-+(1+\eta^2)dx^2-(1+\eta^2)^2x^2(dx^+)^2\,.
\ee
With appropriate rescaling of the light-cone coordinates $x^\pm$ and of $x$ we may absorb the $\eta$ dependence yielding
\be
ds^2=2dx^+dx^-+dx^2-x^2(dx^+)^2\, ,
\ee
corresponding to the plane wave limit of the PCM for the $\nicefrac{SU(2)}{U(1)}$ times a timelike coordinate.

\end{document}